\documentclass[%
reprint,
superscriptaddress,
nofootinbib,
amsmath,amssymb,
aps,
prc,
]{revtex4-1}

\usepackage{siunitx}
\usepackage{graphicx}
\usepackage{dcolumn}
\usepackage{bm}
\usepackage{todonotes}
\usepackage{simplewick}
\usepackage{hyperref}
\usepackage{physics}
\usepackage{verbatim}
\usepackage{float}
\usepackage{array}
\usepackage{lipsum}
\usepackage{svg}
\usepackage{booktabs}
\hypersetup{
     colorlinks,
     linkcolor={blue!50!black},
     citecolor={blue!50!black},
     urlcolor={blue!80!black}
}
\graphicspath{{./figures/}}

\usepackage{bbm}
\usepackage{siunitx}
\usepackage{array}
\usepackage{xspace}
\usepackage{multirow}
\usepackage{tikz}
\usetikzlibrary{shapes.geometric,shapes.misc}
\makeatletter
\DeclareRobustCommand\onedot{\futurelet\@let@token\@onedot}
\def\@onedot{\ifx\@let@token.\else.\null\fi\xspace}

\makeatother

\newcommand{\like}{\mathcal{L}}

\newcolumntype{M}[1]{>{\centering\arraybackslash}m{#1}}

\begin{document}

\title{Global Ab initio Neutrino Mass Limits from Neutrinoless Double-Beta Decay}

\author{T.~Shickele}%
\affiliation{TRIUMF, 4004 Wesbrook Mall, Vancouver, BC V6T 2A3, Canada}%
\affiliation{Department of Physics \& Astronomy, University of British Columbia, Vancouver, BC V6T 1Z1, Canada}

\author{L.~Jokiniemi}%
\affiliation{Technische Universit\"at Darmstadt, Department of Physics, D-64289 Darmstadt, Germany}%
\affiliation{ExtreMe Matter Institute EMMI, GSI Helmholtzzentrum f\"ur Schwerionenforschung GmbH, D-64291 Darmstadt, Germany}%
\affiliation{TRIUMF, 4004 Wesbrook Mall, Vancouver, BC V6T 2A3, Canada}%

\author{A.~Belley}%
\affiliation{TRIUMF, 4004 Wesbrook Mall, Vancouver, BC V6T 2A3, Canada}%
\affiliation{Department of Physics \& Astronomy, University of British Columbia, Vancouver, BC V6T 1Z1, Canada}
\affiliation{Massachusetts Institute of Technology,
Cambridge, Massachusetts 02139, USA}%

\author{J.~D.~Holt}%
\affiliation{TRIUMF, 4004 Wesbrook Mall, Vancouver, BC V6T 2A3, Canada}%
\affiliation{Department of Physics, McGill University, 3600 Rue University, Montr\'eal, QC H3A 2T8, Canada}%

\begin{abstract}

We present global limits for Majorana neutrino masses by combining latest results from neutrinoless double-beta ($0\nu\beta\beta$) decay searches and ab initio nuclear theory. 
Limits are derived in a Bayesian framework utilizing likelihood functions from a suite of $0\nu\beta\beta$-decay experiments in conjunction with nuclear matrix elements calculated from nuclear and electroweak forces derived from chiral effective field theory and implemented in the in-medium similarity renormalization group many-body approach.
In contrast to nuclear models, ab initio results indicate that the current generation of $0\nu\beta\beta$-decay experiments have likely \textit{not} yet reached sensitivities required to probe the mass regime allowed by neutrino-oscillation data, where the combined bounds are notably stronger than those given by individual experiments. 
Finally, from predicted sensitivities of next-generation searches, we show that, while no one individual experiment fully covers the inverted mass ordering, this can be achieved from combined contributions from the four key isotopes: $^{76}$Ge, $^{100}$Mo, $^{130}$Te, and $^{136}$Xe.

\end{abstract}

\maketitle

\section*{Introduction}

Neutrinoless double-beta ($0\nu\beta\beta$) decay is a hypothetical beyond-Standard-Model (BSM) nuclear process in which two neutrons simultaneously undergo $\beta$ decay, without emitting any accompanying antineutrinos. 
This process violates the lepton-number conservation predicted by the Standard Model (SM) of particle physics, and aids in explaining the observed matter-antimatter asymmetry in the Universe~\cite{Fukugita1986,Davidson2008,Agostini2023}. 
Observation of $0\nu\beta\beta$-decay would also reveal that neutrinos are their own antiparticles, or Majorana fermions, and shed light on the neutrino mass mechanism~\cite{Avignone2008,Vergados2012,Ejiri2019}. 

Assuming that $0\nu\beta\beta$-decay is mediated by the exchange of standard light neutrinos\footnote{This commonly studied scenario introduces no new particles, but other exotic mechanisms could also mediate $0\nu\beta\beta$-decay~\cite{MasterFormula,Graf2022,Agostini-Deppisch2023}}, an experimental signal would also illuminate the absolute scale of neutrino masses, which is currently only constrained by neutrino-oscillation experiments~\cite{Esteban2024,SKT2K2025,IceCube2025,NOvA2024,MINOS2020,KamLAND2013}, $\beta$-decay experiments~\cite{KATRIN2025,project8}, and cosmology~\cite{Planck2018}. 
Furthermore, neutrino oscillation results only constrain only the absolute value of the neutrino squared-mass difference $\Delta m_{32}^2 = m_3^2 - m_2^2$~\cite{PDG2024}, leading to either the normal ($m_1<m_2<m_3$) or inverted ($m_3<m_1<m_2$) mass ordering.
Significant effort by several next-generation $0\nu\beta\beta$-decay experiments to fully probe the inverted mass ordering is underway~\cite{Agostini2023}, with many experiments expected to begin data collection within the next decade~\cite{nEXO2022,LEGEND2021,CUPID2024,SNO+2021,AMoRE2025,NEXT2021,PandaX-xT2024,DARWIN2020,XLZD2025}.

To extract the Majorana mass from a given experiment, we need accurate values of the nuclear matrix elements (NME), $M^{0\nu}$, which govern the rate of the decay from the nuclear structure perspective, are isotope-dependent, and can only be provided from nuclear theory. 
Traditionally, the NMEs have been computed using variations of phenomenological nuclear models such the quasiparticle random-phase approximation (QRPA)~\cite{Simkovic2013,Hyvarinen2015,Simkovic2018,Fang2018,Jokiniemi2021}, the nuclear shell model (NSM)~\cite{Horoi2013,Horoi2016,Coraggio2020,Coraggio2022,Jokiniemi2021}, the interacting boson model (IBM)~\cite{Barea2015,Deppisch2020,Kauppinen2025}, density functional theory (DFT)~\cite{Rodriguez2010,LopezVaquero2013,Mustonen2013,Yao2021,Lv2023,Ding2023,Wang2024}, and effective field theory (EFT) \cite{Brase2022}. 
With recent advances in ab initio theory, however, the NMEs can now also be computed from first principles starting only from nuclear and electroweak interactions derived from chiral EFT~\cite{Yao2020,Belley2020,Wirth2021, Novario2020,Yao20light,Yao22DGT,Belley2023HeavyNucleiPreprint,Belley2024}, or hybrid methods such as the NSM combined with quantum Monte Carlo via the generalized contact formalism (GCF)~\cite{Weiss2022}. 
In the past decade, $0\nu\beta\beta$-decay operators have also been derived from an effective-field-theory framework~\cite{Cirigliano2018a,Cirigliano2018,Cirigliano2019,Cirigliano:2020,MasterFormula}, allowing for the first time a consistent ab initio calculation of the NMEs. 
In particular, it was found that to renormalize the theory, a previously unacknowledged contact term had to be promoted to leading order~\cite{Cirigliano2018,Cirigliano2019}, which has been shown to significantly enhance the NMEs~\cite{Wirth2021,Jokiniemi2021,Weiss2022,Belley2023HeavyNucleiPreprint,Belley2024}. 
Other corrections to the traditional operators appear at higher order and generally have a $\lesssim 10\%$ effect on the NMEs~\cite{Pastore2018,Castillo2025}, however a recent study shows that two-body currents, formally appearing at a subleading order, quench NMEs by up to 20\% in light systems~\cite{chambers-wall2025}.

In the present study, we focus on the Valence Space In-Medium Similarity Renormalization Group (VS-IMSRG)---an ab initio method~\cite{Hergert2016,Stroberg2017,Stroberg2019,Miyagi2020} capable of globally accessing most nuclei to the heavy-mass region~\cite{Stro21Drip,Miya22Heavy,Hu22Pb208}, including key candidates of present and future BSM physics searches~\cite{Belley2023HeavyNucleiPreprint, Hu22SDDM,Mart21ISB}---and compute global Bayesian upper limits on the effective Majorana neutrino mass $m_{\beta\beta}$. 
In particular, we follow the approach introduced in Ref.~\cite{Biller2021}, also subsequently used in Refs.~\cite{Jokiniemi2021,Pompa2023}, to derive limits for $m_{\beta\beta}$ by combining  likelihood functions from recent experiments.
In this article we refine these limits with ab initio NMEs, while also deriving combined predicted limits for next-generation experiments. 
For current-generation experiments, results from GERDA~\cite{GERDA2020}, LEGEND-200~\cite{LEGEND2025}, CUPID-Mo~\cite{CUPIDMo2022}, CUORE~\cite{CUORE2024}, EXO-200~\cite{Exo2019} and KamLAND-Zen~\cite{KLZ2025} are used. 
For the global sensitivity reach of next-generation experiments, we consider nEXO~\cite{nEXO2022}, LEGEND-1000~\cite{LEGEND2021}, CUPID and CUPID-1T~\cite{CUPID2024}, SNO+~\cite{SNO+2021}, AMoRE-II~\cite{AMoRE2025}, NEXT-HD~\cite{NEXT2021}, PandaX-xT~\cite{PandaX-xT2024}, DARWIN~\cite{DARWIN2020} and XLZD~\cite{XLZD2025}.

\section*{Theoretical Framework}

\subsection*{Neutrinoless Double-beta Decay}

In this work, we focus only on the standard light-neutrino-exchange mechanism, which allows us to write the decay rate $\Gamma^{0\nu}$ and half-life $T_{1/2}^{0\nu}$ of $0\nu\beta\beta$ decay as
\begin{equation}
 \frac{\Gamma^{0\nu}}{\ln2} = [T_{1/2}^{0\nu}]^{-1} = g_A^4 G^{0\nu} \big( M^{0\nu} \frac{m_{\beta\beta}}{m_e} \big)^2 \;,  
\end{equation}
where $g_{\rm A}$$\sim$1.27 is the unquenched axial-vector coupling constant~\cite{PDG2024,Gysbers2019}, and $G^{0\nu}$ and $M^{0\nu}$ are the phase-space factor~\cite{Kotila2012} and NME, respectively, for each isotope. 
The effective Majorana neutrino mass $m_{\beta\beta} = \sum_i (U_{ei})^2 m_i$, where $m_i$ are the neutrino mass states and $U_{ei}$ are the elements of the PMNS neutrino mixing matrix~\cite{PDG2024}.
The NME is given by
\begin{equation}
    M^{0\nu}=M^{0\nu}_{\rm L} + M^{0\nu}_{\rm S},
\end{equation}
which consists of a long-range (L) part and a recently discovered short-range (S) contribution, required to renormalize the theory in a proper EFT analysis of $0\nu\beta\beta$-decay~\cite{Cirigliano2018,Cirigliano2018a,Cirigliano2019}. We note that $M^{0\nu}_{\rm S}$ is not required for relativistic DFT calculations, such as the MR-CDFT~\cite{Yang2024}.

The long-range NME consists of Gamow-Teller (GT), Fermi (F) and tensor (T) parts:
\begin{equation}
    M^{0\nu}_{\rm L}=M^{0\nu}_{\rm GT}-\left(\frac{g_{\rm V}}{g_{\rm A}}\right)^2M^{0\nu}_{\rm F}+M^{0\nu}_{\rm T}\;,
    \label{eq:long-range_NME}
\end{equation}
where, using the so-called closure approximation, the different terms are evaluated in momentum space as
\begin{equation}
    M^{0\nu}_\alpha = \langle 0^+_f||V_\alpha(\mathbf{q})S_\alpha(\mathbf{q})\tau_1^+\tau_2^+||0^+_i\rangle, \textrm{ $\alpha \in$ [GT,F,T],}
\end{equation}
where $|0^+_i\rangle$ and $|0^+_f\rangle$ are the initial and final nuclear states, $\mathbf{q}$ is the momentum exchange, $\tau_n^+$ is the isospin operator transforming a neutron into a proton, $S_{\alpha}$ are the spin operators 
\begin{align}
    &S_{\rm F} = 1,\\
    &S_{\rm GT} = \boldsymbol{\sigma}_1\cdot\boldsymbol{\sigma}_2,\\
    &S_{\rm T} = -3[(\boldsymbol{\sigma}_1\cdot \hat{q})(\boldsymbol{\sigma}_2 \cdot \hat{q})]+(\boldsymbol{\sigma}_1\cdot\boldsymbol{\sigma}_2)
 \end{align}
with $\hat{q}=\mathbf{q}/||\mathbf{q}||$.
$V_\alpha$ are neutrino potentials given by
\begin{align}\label{eq:4.1-Potential}
    V_\alpha(q) = \frac{R}{2\pi^2}\frac{h_\alpha(q^2)}{|\mathbf{q}|(|\mathbf{q}|+\bar{E})},
\end{align}
where $R=R_0A^{1/3}$ with $R_0=1.2$ fm is the empirical nuclear radius, $\bar{E}$ is the so-called closure energy that approximates the average excitation energy in the intermediate nuclei. 
Different studies in literature make different choices for the closure energy, and we refer the reader to check the original references for the exact values. 
We note that in the pnQRPA approach, one does not use closure energy but instead sums explicitly over the intermediate states.
The forms of the functions $h_{\alpha}$, mainly consisting of nuclear form factors, can be found, e.g., in Refs.~\cite{Engel2016, Agostini2023}.

The short-range NME is given as~\cite{Cirigliano2018,Cirigliano2019,Cirigliano:2020}
\begin{align}
\label{eq:short-range_NME}
    M^{0\nu}_{\rm S} &= \langle 0_f^+||2\frac{g_{\nu}^{\rm NN}}{g_A}\frac{R}{8\pi^3}h_{\rm S}(p,p')\tau^+_1\tau^+_2||0^+_i\rangle,
\end{align}
with a non-locally regulated
contact interaction
\begin{equation}
    h_{\rm S}(p,p')=\left(\frac{m_N g_A^2}{4 f_\pi^2}\right)^2\exp(-\left(\frac{p}{\Lambda}\right)^{2n})\exp(-\left(\frac{p'}{\Lambda}\right)^{2n})
    \label{eq:short-range-potential}
\end{equation}
written in terms of the incoming and outgoing momenta $p$ and $p'$, and the pion decay constant $f_\pi = 92.2$ MeV. 
The coupling $g_{\nu}^{\rm NN}$ is an unknown low-energy constant that can be approximated by charge-independence breaking (CIB) induced by electromagnetism~\cite{Cirigliano2019} or, in an ab initio framework, by fitting it to synthetic data~\cite{Cirigliano:2020,Wirth2021}. 
The regulator power $n$ and cutoff $\Lambda$ in Eq.~\eqref{eq:short-range-potential} are chosen consistently with the nuclear interaction used in the matching, though we note that other studies in the literature \cite{Jokiniemi2021,Weiss2022,Kauppinen2025} use local Gaussian regulators of the form
\begin{equation}
    h_{\rm S}(p,p')\propto \exp(-\frac{(p-p')^2}{2\Lambda^2})\;.
    \label{eq:Gaussian}
\end{equation}

In the present study, we consider only the NMEs that include $M^{0\nu}_{\rm S}$ \eqref{eq:short-range_NME}, which we note represents only a subset of phenomenological results available. 
In particular, we focus on the ab initio NMEs evaluated with the VS-IMSRG method that is capable of accessing all $\beta\beta$ emitters of interest for the current and next-generation experiments: $^{76}$Ge, $^{100}$Mo, $^{130}$Te, and $^{136}$Xe, where we take the corresponding NMEs from Refs.~\cite{Belley2020,Belley2024,Belley2023HeavyNucleiPreprint,AntoineThesis}. 
These NMEs include a non-locally regulated contact term \eqref{eq:short-range-potential} with the coupling adjusted to synthetic data. The regulator power and cutoff is matched to a given chiral interaction used in the studies (see \cite{Belley2020,Belley2024,Belley2023HeavyNucleiPreprint,AntoineThesis} and references therein).
For comparison, we also include most recently available NMEs evaluated with phenomenological methods: the IBM~\cite{Kauppinen2025}, the pnQRPA~\cite{Jokiniemi2021}, the NSM~\cite{Jokiniemi2021}, the MR-CDFT~\cite{Ding2023}, and the hybrid method GCF~\cite{Weiss2022}, where all NMEs are listed in Table \ref{tab:nme_values}. 
These NMEs, excluding the MR-CDFT, include the contact term with the coupling estimated by the CIB of different chiral Hamiltonians with local Gaussian regulators \eqref{eq:Gaussian}. 
In these studies, it has been found that the uncertainty of the contact term mainly originates from the coupling $g_{\nu}^{\rm NN}$, while the regulator-scale dependence is relatively smaller.

\begin{table}[]
\caption{Range of NMEs utilized in the study. Phenomenological NMEs are taken from  IBM~\cite{Kauppinen2025}, pnQRPA~\cite{Jokiniemi2021}, NSM~\cite{Jokiniemi2021}, MR-CDFT~\cite{Ding2023} and GCF~\cite{Weiss2022} calculations. 
Ab initio NMEs are from VS-IMSRG calculations in Refs.~\cite{Belley2020,Belley2023HeavyNucleiPreprint,AntoineThesis}, with a full uncertainty quantification for $^{76}$Ge~\cite{Belley2024} denoted IMSRG(UQ). 
All NMEs include the short-range contribution $M^{0\nu}_{\rm S}$, except for the MR-CDFT values (see text).}
    \label{tab:nme_values}
    \centering
    \begin{ruledtabular}
    \begin{tabular}{lcccc}
         & $^{76}$Ge 
         & $^{100}$Mo
         & $^{130}$Te
         & $^{136}$Xe\\
         \hline & \\[-2.2ex]

         IBM
         & 6.89--8.49 
         & 5.77--7.61 %
         & 4.59--5.82 %
         & 3.72--4.68 \\ %

         pnQRPA 
         & 6.21--9.02 
         & 5.06--8.21 
         & 4.64--6.94 
         & 3.29--4.75 \\

         NSM 
         & 3.86--5.03
         & ---
         & 3.83--5.12 
         & 3.07--4.10 \\

         MR-CDFT  
         & 2.37--6.34
         & 6.01--9.40
         & 2.89--6.02
         & 2.27--5.06 \\
         
         GCF  
         & 2.43--3.79 
         & ---
         & 2.51--3.73 
         & 1.98--2.91 \\
         
         \hline & \\[-2.2ex]
         VS-IMSRG
         & 2.09--3.09
         & 1.80--3.96 
         & 1.52--2.40
         & 1.08--1.90 \\[0.25ex]

         IMSRG (UQ)
         & $2.76_{-1.44}^{+1.26}$
         & --- 
         & ---
         & --- \\

    \end{tabular}
    \end{ruledtabular}
\end{table}

\subsection*{Statistical Methods}

To derive $90\%$ credible interval (CI) global limits and sensitivities on $m_{\beta\beta}$, we consider several different uninformative priors and experimental likelihood functions. 
Paralleling previous Bayesian analyses~\cite{CUPIDMo2022,CUORE2024,GERDA2020,Biller2021,KLZ2025,Zhang2016,LEGEND2025}, priors uniform in $m_{\beta\beta}$, $\Gamma^{0\nu}$ (or equivalently, $m_{\beta\beta}^2$), and $\log(m_{\beta\beta})$ are chosen and compared in our analysis.
The priors are defined on the range $m_{\beta\beta} \in [0,700]$ meV to fully cover the range of values in which the experimental likelihoods are defined (excluding CUPID-Mo), using the most conservative NMEs from the VS-IMSRG. 
Furthermore, the upper limit of 700 meV also roughly corresponds to the most stringent upper limit on $m_{\beta\beta}$ from a past experiment not considered in this work~\cite{HeidelbergMoscow2001}. 
However, for the CUPID-Mo experiment, limits can be less stringent than $m_{\beta\beta} < 700$ meV depending on the choice of NME. 
As such, we only consider a prior uniform in $\Gamma^{0\nu}$ on the range $m_{\beta\beta} \in [0,1500]$ meV in this case. 
For both upper bound choices, variation by $100$ meV leads to differences of 1-4\% in the final limits, which are most pronounced in the case of the $\log(m_{\beta\beta})$ prior.

The relative likelihoods ($\like_{\rm R}$) for the assumed $0\nu\beta\beta$-decay rate based on different experiments are re-parameterized to be in terms of $m_{\beta\beta}$ utilizing the isotope-dependent NMEs given in Table~\ref{tab:nme_values}. 
Then, the combined global relative likelihood can be computed as
\begin{equation}
    \like_{\rm R, Comb} = \prod_{i}^n \like_{{\rm R},i},
\end{equation}
where we multiply the likelihood functions of the $n$ experiments considered. 
From Bayes' Theorem, the posterior distribution for $m_{\beta\beta}$, given the data, is
\begin{equation}
    \mathrm{P}(m_{\beta\beta} \mid \mathrm{Data}) \propto 
    \mathcal{L}_{\mathrm{Comb}}(\mathrm{Data} \mid m_{\beta\beta}) ~ 
    \pi(m_{\beta\beta}),
\end{equation}
where $\mathcal{L}_{\mathrm{Comb}}$ denotes the combined likelihood and 
$\pi(m_{\beta\beta})$ the chosen prior on the effective Majorana mass. 
Integrating this posterior over positive $m_{\beta\beta}$ allows us to obtain the $90\%$ CI global limits from current experiments, and sensitivities of next-generation experiments.

\subsection*{Experimental Inputs}

\begin{figure}[h!]
    \centering
    \includegraphics[width=\linewidth]{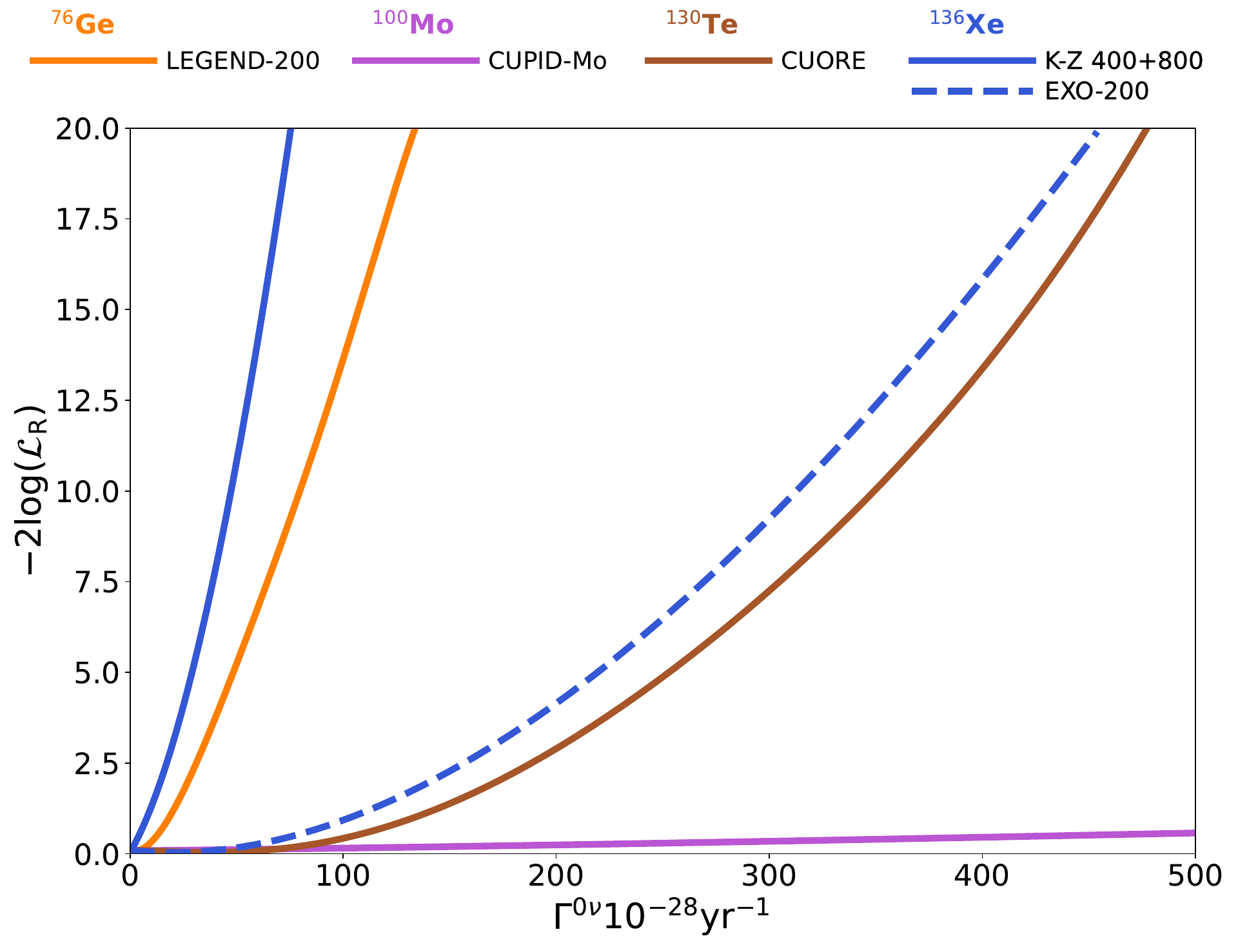}
    \caption{The likelihood functions for current and recently completed experiments, LEGEND-200~\cite{LEGEND2025} (combined with results from GERDA~\cite{GERDA2020} and MAJORANA~\cite{MAJORANA2023}), CUPID-Mo~\cite{CUPIDMo2022}, CUORE~\cite{CUORE2024}, EXO-200~\cite{Exo2019}, and KamLAND-Zen (including both KamLAND-Zen 400 and 800)~\cite{KLZ2025}.}
    \label{fig:current_likelihoods}
\end{figure}

\begin{figure}[h!]
    \centering
    \includegraphics[width=\linewidth]{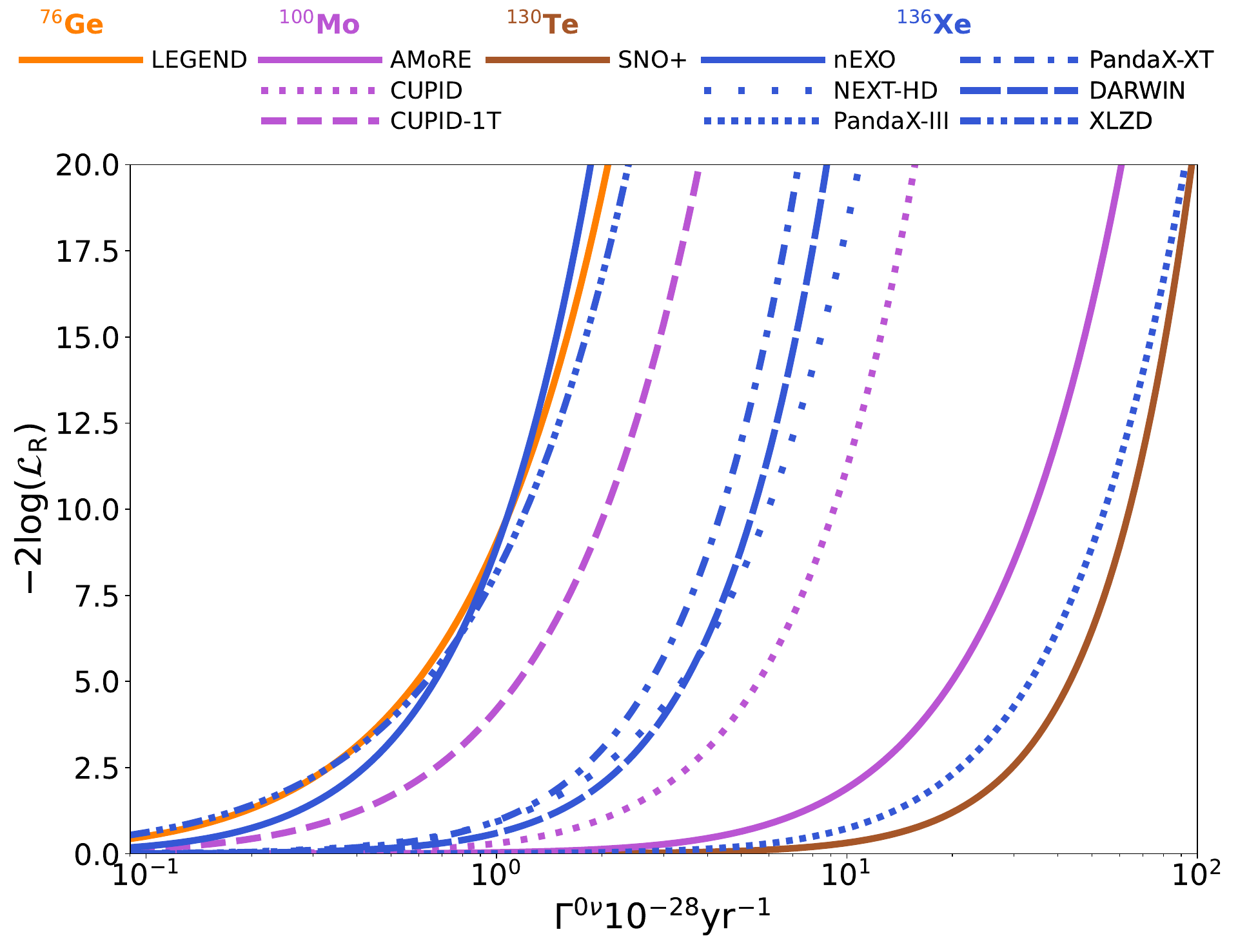}
    \caption{The likelihood functions for the next-generation experiments, LEGEND-1000~\cite{LEGEND2021}, AMoRE~\cite{AMoRE2025}, CUPID~\cite{CUPID2024}, CUPID-1T~\cite{CUPID2024}, SNO+~\cite{SNO+2021}, nEXO~\cite{nEXO2022}, and NEXT-HD~\cite{NEXT2021}. Note the x-axis is shown in log scale, unlike Fig.~\ref{fig:current_likelihoods}.}
    \label{fig:nextgen_likelihoods}
\end{figure}

Here, we present the likelihood functions utilized for current and next-generation $0\nu\beta\beta$-decay experiments. 
We utilize the likelihoods available in the literature when possible, and otherwise employ a Poisson counting experiment model~\cite{FeldmanandCousins1998,Rolke2005,Cowan2011} to determine our likelihood functions.
Negative log-likelihoods are shown for the current and next-generation experiments in Fig.~\ref{fig:current_likelihoods},~\ref{fig:nextgen_likelihoods}.

The likelihood function for MAJORANA~\cite{MAJORANA2023}, GERDA~\cite{GERDA2020} and LEGEND-200~\cite{LEGEND2025} is obtained from the posterior distribution of a combined analysis of the three experiments in Ref.~\cite{LEGEND2025}. 
As the results utilize $\Gamma^{0\nu}$ uniform prior, the posterior directly corresponds to the required likelihood function. 
As expected, we find the derived $90\%$ CI $0\nu\beta\beta$-decay half-life limit in $^{76}$Ge to agree with the reported Bayesian limit of $T_{1/2}^{0\nu} > 1.9 \times 10^{26}$ yr.

Similarly, the CUPID-Mo likelihood function utilizes the posterior distribution reported by the CUPID-Mo collaboration in Ref.~\cite{CUPIDMo2022}. 
Their results also conveniently assume a prior uniform in $\Gamma^{0\nu}$, and the reconstructed half-life limit matches the expected limit of $T_{1/2} > 1.8 \times 10^{24}$ yr obtained in the reported results.

The likelihood function for EXO-200 Phases I and II~\cite{Exo2019} is modeled based on a Poisson distribution, with the addition of marginalization over the background uncertainty to account for systematic errors~\cite{Cowan1998}.
Using a toy Monte Carlo, this results in a slightly larger half life than the limit reported by the collaboration ($4.3 \times 10^{25}$ yr compared to $3.5 \times 10^{25}$ yr at $90\%$ CI). 
This difference may be caused by the different statistical approaches (frequentist vs. Bayesian) and the detailed event analysis performed by the EXO-200 collaboration, over the simple counting analysis employed here.
However, this matches the Bayesian result obtained previously in Ref.~\cite{Biller2021} and fortunately, the difference is negligible in influencing our final combined constraints on $m_{\beta\beta}$.

For the CUORE likelihood, we utilize the posterior distribution obtained by the CUORE collaboration as was done for the $^{76}$Ge experiments and CUPID-Mo. We note a prior uniform in $\Gamma^{0\nu}$ is assumed in their Bayesian analysis~\cite{CUORE2024}. Using this posterior as a likelihood leads to a half-limit of $T_{1/2} > 3.8 \times 10^{25}$ yr, matching the published result.

The likelihood function combining KamLAND-Zen 400 and 800 is obtained from the $\Delta\chi^2$ profile in Ref.~\cite{KLZ2025,ItaruShimizu} through use of Wilks' theorem~\cite{Wilks1938}. 
However, the possibility that $0\nu\beta\beta$-decay does not exist such that the true signal rate is zero necessitates the consideration of Chernoff's extension~\cite{Chernoff1954,Algeri2020,SelfLiang1987}. 
Comparing the half-life limits obtained with and without Chernoff's extension reveals that in both cases, our Bayesian limits are more conservative than reported by the KamLAND-Zen collaboration~\cite{KLZ2025}. 
Therefore, we disregard Chernoff's extension and obtain a limit of $T_{1/2} > 3.0 \times 10^{26}$ yr, to better reproduce the reported sensitivity of the KamLAND-Zen experiment ($T_{1/2} > 3.8 \times 10^{26}$ yr). 
Inclusion of Chernoff's extension when only considering current-generation limits would weaken combined limits by $5-15$ meV, depending on the choice of NMEs, which, while non negligible, does not alter the conclusions of this work.

Projected likelihood functions to estimate the sensitivity reach of next-generation experiments are derived based on a Poisson distribution. We sample the sensitivity curves obtained from the Poisson processes to ensure results match the median sensitivity estimates in each experiment. Here, we focus on the nEXO~\cite{nEXO2022}, LEGEND-1000~\cite{LEGEND2021}, CUPID and CUPID-1T~\cite{CUPID2024}, SNO+~\cite{SNO+2021}, AMoRE-II~\cite{AMoRE2025}, NEXT-HD~\cite{NEXT2021}, PandaX-xT~\cite{PandaX-xT2024}, DARWIN~\cite{DARWIN2020} and XLZD (80T scenario)~\cite{XLZD2025} next-generation experiments, which all aim to partially or fully probe the inverted mass ordering in the coming years.

\section*{Results and Discussion}

The $90\%$ CI upper limits on $m_{\beta\beta}$ for each isotope and their combination from currently running and recently completed experiments is shown in Fig.~\ref{fig:current_limits} (see Tables \ref{tab:indiv_limits_current} and \ref{tab:indiv_limits_future} in Appendix for detailed numbers). 
The allowable regions of $m_{\beta\beta}$ in both the inverted and normal mass orderings of the neutrino masses are included for comparison. 
The darker regions of the mass hierarchies indicate areas of the phase space allowed at $1 \sigma$ from neutrino-oscillation experiments, while lighter bands indicate the $3\sigma$ region. 
Limits obtained from phenomenological NMEs are shown in gray, in comparison to results using ab initio VS-IMSRG NMEs. 
We show the limits obtained for each isotope separately: $^{76}$Ge bands use the likelihood function from the combination of MAJORANA, GERDA and LEGEND-200, $^{100}$Mo use CUPID-Mo, $^{130}$Te bands correspond to CUORE, and $^{136}$Xe bands combine EXO-200 and KamLAND-Zen 800. 
The light blue bands show the upper limits obtained from combining all experimental results. 
While we see the upper limits derived from phenomenological NMEs suggest that xenon-based experiments have already partially probed the inverted mass ordering, ab initio results strongly suggest that this has not yet been accomplished, either individually or collectively. 
Note that here we consider an unquenched axial-vector coupling $g_{\rm A}=1.27$, which often results in an overprediction of $\beta$ and standard $\beta\beta$ decays in phenomenological frameworks, necessitating corrections known as ``$g_{\rm A}$ quenching". 
However, the situation is less clear for $0\nu\beta\beta$ decay operating at much higher momentum exchange regime \cite{Engel2017}. 
To obtain more conservative limits, one could use for example a moderately quenched effective coupling $g_{\rm A}^{\rm eff}=1.0$, which would reduce the long-range NMEs \eqref{eq:long-range_NME} by $20\%-30\%$ and shift the $m_{\beta\beta}$ limits some $25\%-50\%$ higher \cite{Jokiniemi2021}.

For each isotope, we assume the NMEs to be uniformly distributed within their range, leading to the aforementioned upper-limit bands. 
The phenomenological bands combine the limits obtained separately for each method: IBM, pnQRPA, NSM, MR-CDFT, and GCF NME (see the values in Table~\ref{tab:nme_values}).
The ab initio bands correspond to the spread of values obtained from the VS-IMSRG~\cite{Belley2023HeavyNucleiPreprint} using different nuclear interactions derived from chiral EFT. 
We note that there are correlations between the NMEs of different isotopes:~each individual interaction tends to give comparatively smaller or larger values across the isotopes. Additionally, for $^{76}$Ge, we present the additional result from a Bayesian analysis of NME errors in the IMSRG(UQ)~\cite{Belley2024}, illustrated as a vertical distribution of limits. 
Here, the $68\%$ credible interval of this final limit distribution is highlighted in colour, while we indicate the limits obtained by a direct conversion of the interval reported in Ref.~\cite{Belley2024} as a dark gray error bar. 
The difference between the two intervals is due to the non-linear transformation between the NMEs and limits, which amplifies smaller values of the limit distribution.
For consistency in the final combined limit, we use the uniformly distributed ranges for all isotopes, but work is currently in progress to obtain quantified uncertainties for all isotopes of interest, which will allow us to study the correlations between different isotopes more carefully. 
Consequently, the present NME bands should be viewed as coverage ranges rather than statistically consistent uncertainty intervals.

\begin{figure}[t]
    \centering
    \includegraphics[width=\linewidth]{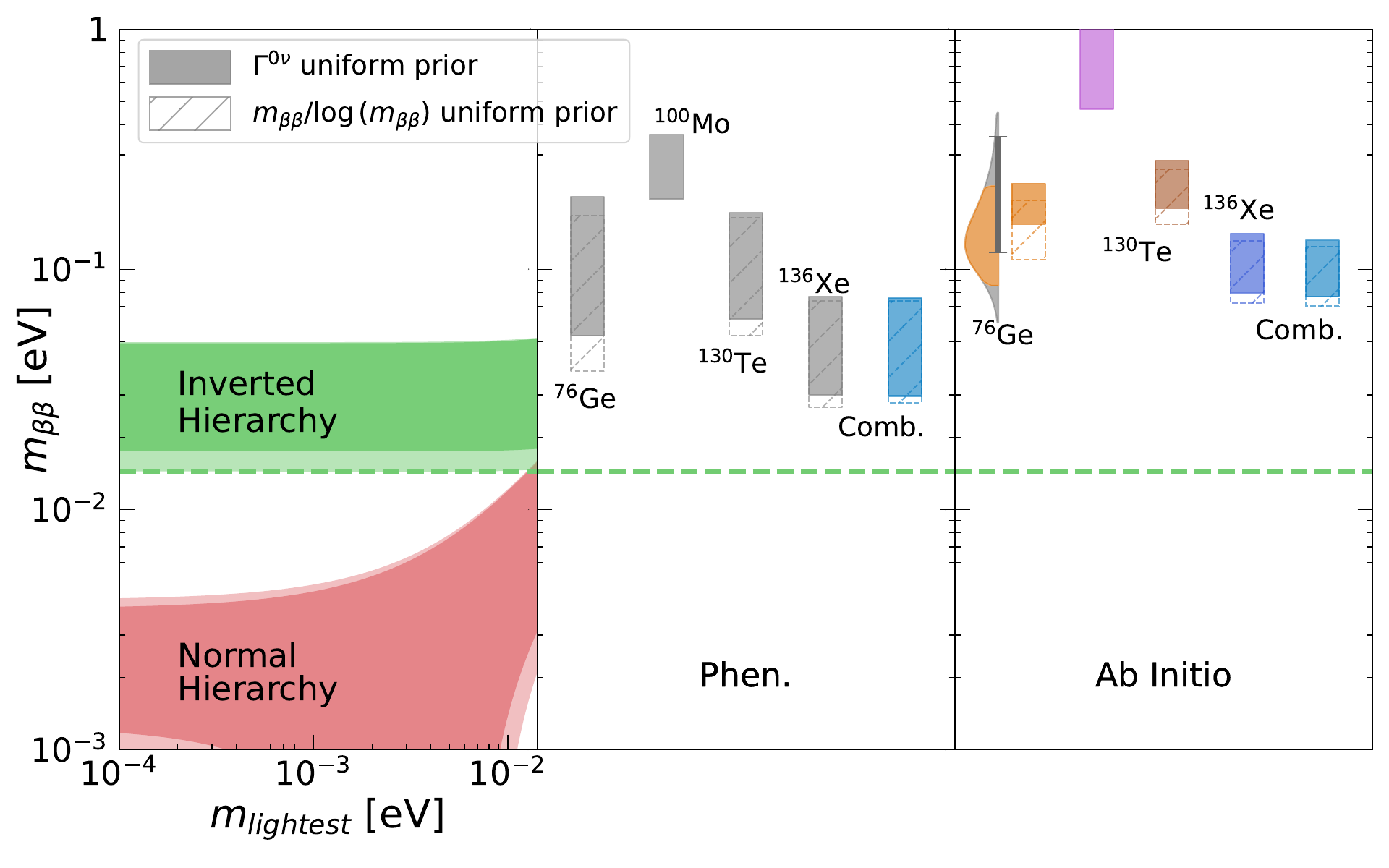}
    \caption{Limits on $m_{\beta\beta}$ from current experiments~\cite{MAJORANA2023,GERDA2020,LEGEND2025,CUPIDMo2022,CUORE2024,Exo2019,KLZ2025} and their combination.
    Theory bands correspond to the range of NMEs, where
    $^{76}$Ge results are given as a distribution using the posterior in Ref.~\cite{Belley2024} (68\% interval in colour). 
    Solid bars indicate limits obtained with a $\Gamma^{0\nu}$ uniform prior, while hatched bars indicate the increased reach obtained when using priors uniform in $m_{\beta\beta}$ and $\log(m_{\beta\beta})$. }
    \label{fig:current_limits}
\end{figure}

In Fig.~\ref{fig:current_limits} we show results obtained with different priors on $m_{\beta\beta}$, where limits obtained from a $\Gamma^{0\nu}$ (or, equivalently, $m_{\beta\beta}^2$) uniform prior (shown in the solid bars) are the most conservative and match previous results~\cite{CUPIDMo2022,CUORE2024,GERDA2020,Biller2021,KLZ2025,Zhang2016,LEGEND2025}. 
The combined global limits in this case are $m_{\beta\beta} \le 30-76$~meV and using the phenomenological NMEs and $m_{\beta\beta} \! \le \! 77-132$ meV from ab initio. 
Limits using priors uniform in $m_{\beta\beta}$ and $\log(m_{\beta\beta})$ are also included as hatched bars, where $m_{\beta\beta}$ uniform priors result in $\gtrsim5\%$ more stringent limits than those using $\Gamma^{0\nu}$ uniform priors. 
The $\log(m_{\beta\beta})$ uniform priors result in the most stringent limits, reaching $m_{\beta\beta} \le 28-72$ meV with phenomenological NMEs and $m_{\beta\beta} \le 70-118$ meV with ab initio, where for both sets, the strongest current limit arises from KamLAND-Zen, regardless of the choice of prior. 
In the case of ab initio NMEs with a $\Gamma^{0\nu}$ uniform prior, the combined limits are $3-8$ meV more stringent than the individual KamLAND-Zen bound\footnote{Since the reconstructed likelihood leads to a limit less stringent than that obtained by Ref.~\cite{KLZ2025}, the combined limit is comparable to directly converting those results into a limit on $m_{\beta\beta}$}.

\begin{figure}[t]
    \centering
    \includegraphics[width=\linewidth]{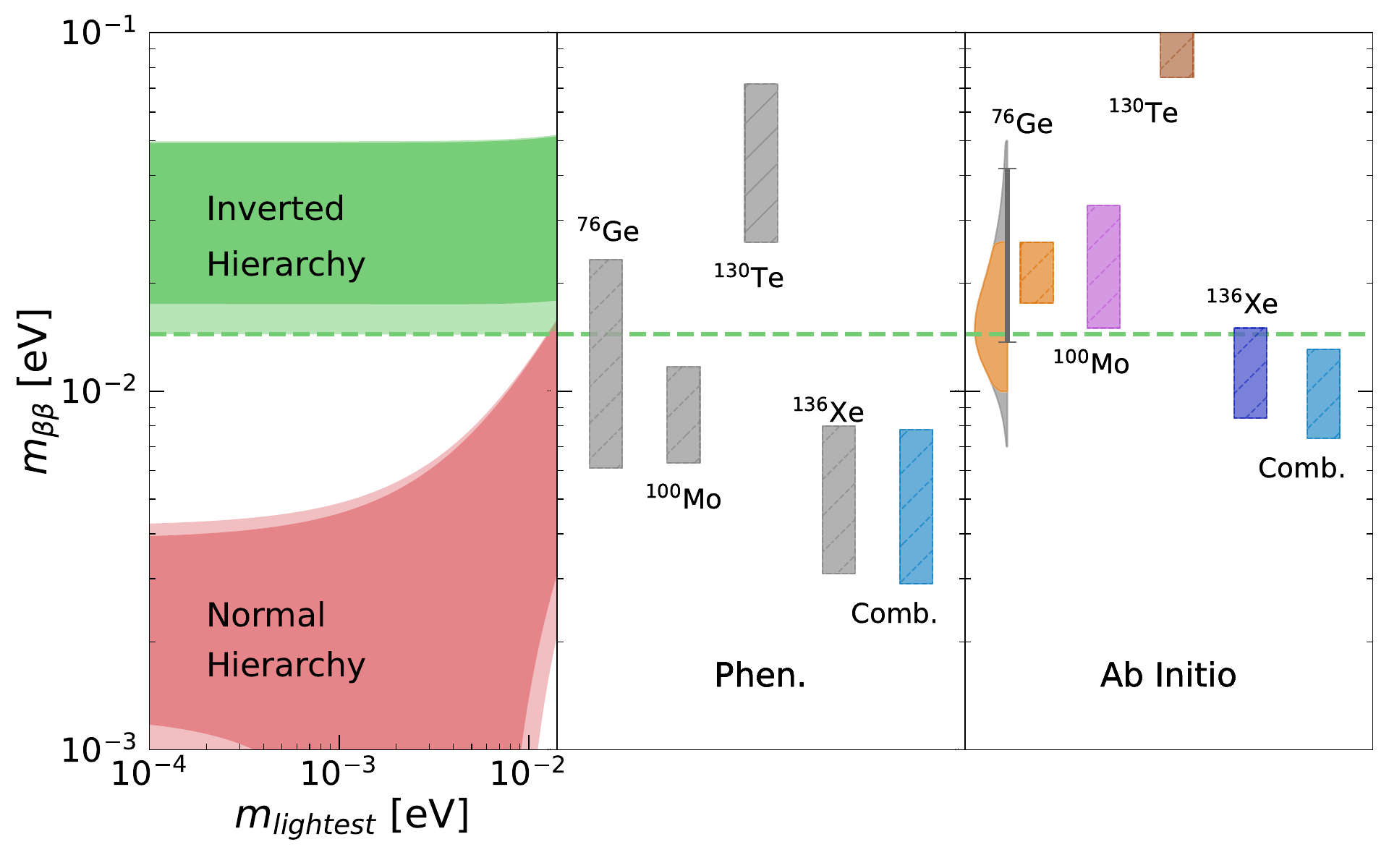}
    \caption{Next-generation $m_{\beta\beta}$ sensitivities from worldwide planned searches, where ab initio projections for LEGEND-1000~\cite{LEGEND2021} utilize the NME range obtained from Ref.~\cite{Belley2020} and posterior from Ref.~\cite{Belley2024}. 
    The $^{100}$Mo projections are obtained from the combination of CUPID~\cite{CUPID2024} and AMoRE-II~\cite{AMoRE2025}, and $^{130}$Te from SNO+~\cite{SNO+2021}. 
    For $^{136}$Xe we combine sensitivities from nEXO~\cite{nEXO2022}, NEXT-HD~\cite{NEXT2021}, PandaX-xT~\cite{PandaX-xT2024}, DARWIN~\cite{DARWIN2020} and XLZD~\cite{XLZD2025}. }
    \label{fig:nextgen_limits}
\end{figure}

Fig.~\ref{fig:nextgen_limits} presents the projected $90\%$ exclusion sensitivities of next-generation $0\nu\beta\beta$-decay experiments around the world, and their combined reach. 
Sensitivity projections for LEGEND-1000 are used for $^{76}$Ge, the combination of CUPID and AMoRE-II for $^{100}$Mo, SNO+ for $^{130}$Te, and the combination of nEXO, NEXT-HD, PandaX-xT and DARWIN for $^{136}$Xe. 
While we do not include the effects of CUPID-1T, the proposed successor to CUPID utilizing $^{100}$Mo, we nevertheless show such results in the Appendix for reference.

With phenomenological NMEs, the upper limits obtained suggest that several experiments, including nEXO and CUPID, may fully probe the inverted mass ordering for the range of NMEs including $M^{0\nu}_{\rm S}$. 
The global $90\%$ exclusion sensitivity based on the phenomenological NMEs is between $m_{\beta\beta} \le 2.9-7.8$ meV, and $m_{\beta\beta} \le 2.3-4.1$ meV with CUPID-1T included. 
On the other hand, with ab initio NMEs, we find that no individual next-generation experiment will fully probe the inverted mass ordering with confidence, but this can be achieved as long as one combines the proposed reach of next-generation experiments, providing a final combined reach of $m_{\beta\beta} \le 7.4-13.1$ meV, or $m_{\beta\beta} \le 5.6-10.8$ meV with CUPID-1T considered.
The absence of a single experiment dominating the combined reach also demonstrates that a worldwide effort, with several experiments operating with different isotopes in conjunction is needed to achieve the goal of probing the inverted mass ordering. 
Furthermore, in the event of a successful detection, operation of experiments across several different isotopes would allow for the differentiation between different $0\nu\beta\beta$-decay mechanisms beyond the standard light neutrino exchange mechanism discussed here~\cite{Graf2022}.
Finally, we note that the results for the next-generation sensitivities vary negligibly between different prior choices, which is to be expected, as the improved reach of next-generation experiments results in enough statistical information to overwhelm our prior choice.

\section*{Conclusion}

We have derived global limits for the effective Majorana-neutrino mass $m_{\beta\beta}$ from the sensitivities of current and future $0\nu\beta\beta$-decay experiments combined with current ab initio nuclear matrix elements. 
The results show that, while ab initio methods tend to give less stringent limits for $m_{\beta\beta}$ for each experiment, combining the current limits through the likelihood-function method moves the limit closer to the inverted mass ordering. 
Further, while ab initio matrix elements suggest none of the currently planned next-generation experiments will be capable of fully probing the inverted mass ordering, combining their sensitivity projections shows that the combined global reach on $m_{\beta\beta}$ successfully covers the entire inverted mass ordering. 
Ongoing work aiming at rigorous ab initio uncertainty quantification of NMEs in all relevant isotopes will allow a direct comparison of limit posteriors across experiments, thereby informing future searches in the eventual aim to reach the normal hierarchy. 
Development of machine-learning emulators such as BANNANE~\cite{Belley2026} will facilitate such an extended analysis of the various theoretical uncertainties arising from the chiral interactions, operators, many-body method and more.

\section*{Acknowledgements}
We would like to thank G. Benato, J. Detwiler, M. Drissi, B. Lenardo, J. Men\'{e}ndez, T. Miyagi, and A. Todd for useful discussions. 
We also thank I. Shimizu for providing the KamLAND-Zen likelihood and A. Lindote for aiding in reconstructing the XLZD sensitivity curve.
TRIUMF receives funding via a contribution through the National Research Council of Canada. 
This work was further supported by NSERC under grants SAPIN-2024-0003 and PDF-587464-2024, the Arthur B. McDonald Canadian Astroparticle Physics Research Institute, and the Canadian Institute for Nuclear Physics.
L.J. acknowledges support of the LOEWE Top Professorship LOEWE/4a/519/05.00.002(0014)98 by the State of Hesse and A.B. acknowledges the support of the Nat[PDF-587464-2024].
Computations were performed with an allocation of computing resources on Cedar at WestGrid and the Digital Research Alliance of Canada.

\bibliography{library}

\clearpage

\appendix*

\section{Numerical values for the Majorana-mass limits}

In Tables~\ref{tab:indiv_limits_current} and \ref{tab:indiv_limits_future}, we collect the upper limits on the $m_{\beta\beta}$ presented in Figs.~\ref{fig:current_limits} and \ref{fig:nextgen_limits} in the main text.

\begin{figure}[h!]
    \centering
    \includegraphics[width=\linewidth]{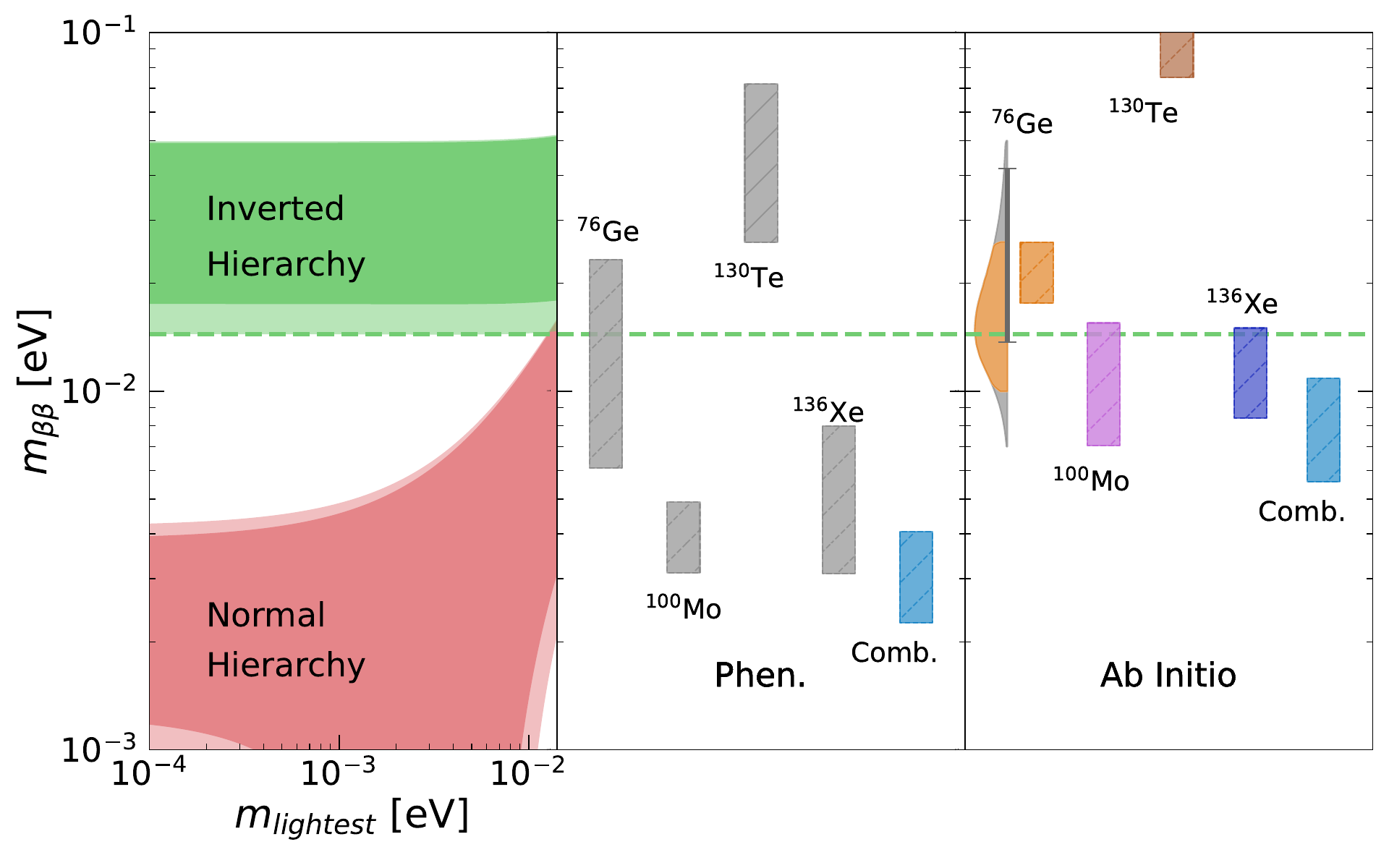}
    \caption{Next-generation $m_{\beta\beta}$ sensitivities from upcoming experiments when including CUPID-1T~\cite{CUPID2024}.}
    \label{fig:nextgen_limits_wCUPID}
\end{figure}

In addition, in Fig.~\ref{fig:nextgen_limits_wCUPID}, we include the sensitivity to $m_{\beta\beta}$ from next-generation experiments when CUPID-1T is also considered. CUPID-1T is shown to significantly improve the sensitivity in $^{100}$Mo and the global combination in the ab initio case. The effect is not as pronounced when using phenomenological NMEs, but nevertheless leads to a improved sensitivity.

\begin{table}[]
\caption{Upper limits on the $m_{\beta\beta}$ (in meV) from current generation $0\nu\beta\beta$-decay experiments.}
    \label{tab:indiv_limits_current}
    \centering
    \begin{ruledtabular}
    \begin{tabular}{lccccc}
         & $^{76}$Ge 
         & $^{100}$Mo
         & $^{130}$Te
         & $^{136}$Xe
         & Global\\
         \hline & \\[-2.2ex] 

         IBM
         & 56--69
         & 242--319
         & 74--94
         & 33--41
         & 31--39\\ 

         pnQRPA 
         & 53--77
         & 224--364
         & 62--93
         & 32--46
         & 30--44 \\

         NSM 
         & 95--123
         & ---
         & 84-113
         & 37--50
         & 37--49 \\
         
         GCF  
         & 126--196 
         & ---
         & 116--172 
         & 52--77
         & 52--76 \\

         MR-CDFT 
         & 75--201
         & 196--306
         & 72--149
         & 30--67
         & 30--66 \\
         \hline & \\[-2.2ex]
         
         VS-IMSRG
         & 154--228
         & 465--1023 
         & 180--284
         & 80--141
         & 77--132 \\[0.25ex]

         VS-IMSRG~\cite{Belley2024}
         & 119--139
         & --- 
         & ---
         & --- 
         & --- \\

    \end{tabular}
    \end{ruledtabular}
\end{table}

\begin{table}[]
\caption{Upper limits on the $m_{\beta\beta}$ (in meV) from next-generation $0\nu\beta\beta$-decay experiments excluding CUPID-1T}
    \label{tab:indiv_limits_future}
    \centering
    \begin{ruledtabular}
    \begin{tabular}{lccccc}
         & $^{76}$Ge 
         & $^{100}$Mo
         & $^{130}$Te
         & $^{136}$Xe
         & Global\\
         \hline & \\[-2.2ex] 

         IBM
         & 6.5--8.0
         & 7.8--10.3
         & 31--39
         & 3.4--4.3
         & 3.1--3.8\\ 

         pnQRPA 
         & 6.1--8.9
         & 7.2--11.7
         & 26--39
         & 3.4--4.8
         & 3.0--4.3 \\

         NSM 
         & 11--14
         & ---
         & 35--47
         & 3.9--5.2
         & 3.8--5.0 \\
         
         GCF  
         & 15--23
         & ---
         & 48--72
         & 5.5--8.0
         & 5.3--7.8 \\

         MR-CDFT  
         & 8.7--23.3
         & 6.3--9.9
         & 30--63
         & 3.1--7.0
         & 2.9--6.1 \\
         \hline & \\[-2.2ex]
         
         VS-IMSRG
         & 18--26
         & 15--33
         & 75--119
         & 8.4--15
         & 7.4--13.1 \\[0.25ex]

         VS-IMSRG~\cite{Belley2024}
         & 14--42
         & --- 
         & ---
         & --- 
         & --- \\

    \end{tabular}
    \end{ruledtabular}
\end{table}

\end{document}